\documentclass[11pt]{article}
\usepackage[dvips]{graphicx,epsfig,color}
\usepackage{wrapfig,rotating}
\usepackage{amssymb,amsmath,array}
\begin{document}
\title{\bf
Slepton NLG (Non-Linear Gauge) in GRACE/SUSY-loop} 
\author{Masato JIMBO$^{1)}$, Tadashi ISHIKAWA$^{2)}$ and Masaaki KURODA$^{3)}$
\vspace{.3cm}\\
{\it
1) Chiba University of Commerce, Ichikawa, Chiba 272-8512, Japan
}
\vspace{.1cm}\\
{\it
2) KEK, Tsukuba, Ibaraki 305-0801, Japan
}
\vspace{.1cm}\\
{\it
3) Meiji Gakuin University, Yokohama, Kanagawa 244-8539, Japan
}
\date{}
}

\maketitle
\thispagestyle{empty}

\begin{abstract}
We have been developing a program package called {\tt GRACE/SUSY-loop} which is
for the automatic calculations of the MSSM amplitudes in one-loop order.
The non-linear gauge (NLG) fixing conditions play the crucial role in
the calculations in one-loop order which contain a large number of Feynman
diagrams.  We present the recent progress in {\tt GRACE/SUSY-loop} which
is obtained by extending the non-linear gauge formalism to the slepton sector.
\end{abstract}

\section{Introduction}

Supersymmetry (SUSY) between bosons and fermions at the unification-energy
scale is one of the most promising hypotheses in the theory beyond the
standard model (BSM), which is expected to resolve the remaining problems
in the standard model (SM).  The minimal supersymmetric extension of the SM
(MSSM) is consistent with all the known high-precision experiments at
a level comparable to the SM.

Since it is a broken symmetry at the electroweak-energy scale, the relic of
SUSY is expected to remain as a rich spectrum of heavy SUSY particles, i.e.
partners of usual matter fermions (leptons and quarks), gauge bosons and
Higgs bosons, which are named sfermions (sleptons and squarks), gauginos
and higgsinos, respectively.  The quest of these new particles is one of
the most important aims of the high-energy physics at present and future
colliders of sub-TeV-energy or TeV-energy region.

In particular, experiments at the ILC offer high-precision determination of
SUSY parameters via $e^- e^+$-annihilation processes.  Since the theoretical
predictions with the high accuracy comparable to that of experiments is required
to extract important physical results from the data, we have to include at least
one-loop contributions in perturbative calculations of amplitudes.

Recently, we have calculated the radiative corrections to production processes
and decay processes of SUSY particles in the framework of the MSSM using
{\tt GRACE/SUSY-loop}~\cite{Fujimoto:2007bn, Iizuka:2010bh, Jimbo:2010} which
is a program package for automated computations of the MSSM in one-loop order.
For the test of numerical results, we have used the non-linear gauge (NLG)
formalism~\cite{Fujikawa:1973, Gavela:1981, Haber:1988, Capdequi:1990, Boudjema:1996, Kato:2006}
applied to {\tt GRACE/SUSY-loop}~\cite{Fujimoto:2006km}.

In this paper, we show the recent progress in {\tt GRACE/SUSY-loop} which is
obtained by extending the non-linear gauge formalism to the slepton sector.

\section{Features of GRACE/SUSY-loop}

For many-body final states, each production process or decay process is described by
a large number of Feynman diagrams even in tree-level order.  There are still more
Feynman diagrams in one-loop order even for two-body final states.  For this reason,
we have developed the {\tt GRACE} system~\cite{Yuasa:1999rg}, which enables us to calculate
amplitudes automatically.

A program package called {\tt GRACE/SUSY-loop} is the version of the {\tt GRACE}
system for the calculation of the MSSM amplitudes in one-loop order, which includes
the model files of the MSSM and can produce corresponding 2-point functions and counter terms.
For the automatic calculation of the MSSM amplitudes in one-loop order,
there exist other program packages independently developed by other groups,
{\tt SloopS}~\cite{Baro:arXiv0906.1665} and {\tt FeynArt/Calc}~\cite{Hahn:2000jm}.

As explained in \cite{Fujimoto:2007bn}, the renormalization scheme adopted for
the electroweak (EW) interactions in {\tt GRACE/SUSY-loop} is a variation of the on-mass-shell
scheme~\cite{Chankowski:1994, Yamada:1991, Dabelstein:1995, Hollik:2002, Fritzsche:2002},
which is an MSSM extension of the scheme in the SM used in {\tt GRACE-loop}~\cite{Yuasa:1999rg}.  
There are some degrees of freedom in the renormalization conditions of the sfermion sector.  We can
choose different sets of residue conditions, decoupling conditions on the transition
terms between the lighter and the heavier sfermions.

In {\tt GRACE/SUSY-loop}, we use the technique of the NLG formalism~\cite{Fujikawa:1973, Gavela:1981,
 Haber:1988, Capdequi:1990, Boudjema:1996, Kato:2006} in order to confirm the validity of calculations
by imposing the NLG invariance on physical results.  The NLG formalism is an extension of
the linear $R_\xi$-gauge.
The gauge fixing lagrangian for the EW interactions in the NLG~\cite{Fujimoto:2006km, Baro:arXiv0906.1665}
is given as follows:
{\small
\begin{eqnarray}
  {\cal L}_{\rm gf} &=&  -{\frac{1}{\xi_W}}\vert F_{W^+} \vert^2 
                      -{\frac{1}{2\xi_Z}}(F_Z)^2
                      -{\frac{1}{2\xi_\gamma}}(F_\gamma)^2, \label{L1}\\
  F_{W^\pm} &=& (\partial_\mu \pm ie\tilde\alpha A_\mu  
                  \pm ig c_W\tilde\beta Z_\mu) W^{\pm\mu}
            \pm i\xi_W {\frac{g}{2}}(v + \tilde\delta_H H^0 + \tilde\delta_h h^0 
              \pm i\tilde\kappa G^0)G^\pm , \label{L2}\\
   F_Z &=& \partial_\mu Z^\mu+\xi_Z{\frac{g_Z}{2}}(v +\tilde\epsilon_H H^0
                +\tilde\epsilon_h h^0)G^0, \label{L3} \\
   F_\gamma &=& \partial_\mu A^\mu, \label{L4}
\end{eqnarray}
}where {\small $ v = \sqrt{v_1^2+v_2^2}$} , {\small $M_W = \frac{gv}{2}$} ,
{\small $ M_Z = \frac{g_Z v}{2}$} , {\small $h^0$} and {\small $H^0$} stands for
the lighter and heavier CP even Higgs boson, respectively, {\small $G^\pm$} and {\small $G^0$}
stands for the Goldstone boson which corresponds to gauge boson {\small $W^\pm$} and
{\small $Z$}, respectively.  The gauge fixing lagrangian (\ref{L1}) contains seven
independent NLG-parameters, ($\tilde\alpha, \tilde\beta, \tilde\delta_H, \tilde\delta_h,
\tilde\kappa, \tilde\epsilon_H, \tilde\epsilon_h$).
The numerical tests are performed by varying these parameters.

\section{Renormalization conditions in the slepton sector}
The bare mass term in the slepton sector of MSSM lagrangian is given by
{\small
\begin{equation}
  {\cal L}_0^{mass} = -
    \begin{pmatrix}
      \tilde \ell_L^* &
      \tilde \ell_R^*
    \end{pmatrix}_0
    \begin{pmatrix}
      m^2_{\tilde \ell_L} & m^2_{\tilde \ell_{LR}} \\ 
\rule{0mm}{11pt}
      m^{2*}_{\tilde \ell_{LR}} & m^2_{\tilde \ell_R}
    \end{pmatrix}_0
    \begin{pmatrix}
      \tilde \ell_L \\
      \tilde \ell_R
    \end{pmatrix}_0
 ~,~~\ell = e, \mu, \tau~, \label{3.1}
\end{equation}
}where
{\small
\begin{eqnarray}
    m^2_{\tilde \ell_L} &=&~ \widetilde m^2_{\tilde \ell_L}+ m^2_\ell 
                        +M_Z^2\cos 2\beta (T_{3\ell}-Q_\ell \sin^2 \theta_W)~, \nonumber \\
    m^2_{\tilde \ell_R} &=&~ \widetilde m^2_{\tilde \ell_R}+ m^2_\ell 
                        +M_Z^2\cos2\beta Q_\ell \sin^2 \theta_W~,\label{3.2} \\
    m^2_{\tilde \ell_{LR}} &=& -m_\ell (A_\ell + \mu \tan\beta) ~.   \nonumber
\end{eqnarray}
}It contains three parameters $\widetilde m_{\ell_L}$, $\widetilde m_{\ell_R}$ and $A_\ell$.
Diagonalizing the mass matrix, we determine the mixing angle
$\theta_\ell$ and the mass eigenvalues $m_{\tilde \ell_1},m_{\tilde \ell_2}$
\footnote{We use the convention, $m_{\tilde \ell_1}<m_{\tilde \ell_2}$ },
{\small
\begin{equation}   
    \begin{pmatrix}
      \cos\theta_\ell & \sin\theta_\ell \\
     -\sin\theta_\ell & \cos\theta_\ell
    \end{pmatrix}_0
    \begin{pmatrix}
      m^2_{\tilde \ell_L} & m^2_{\tilde \ell_{LR}} \\ 
\rule{0mm}{11pt}
      m^{2*}_{\tilde \ell_{LR}} & m^2_{\tilde \ell_R}
    \end{pmatrix}_0
    \begin{pmatrix}
      \cos\theta_\ell & -\sin\theta_\ell \\
      \sin\theta_\ell & \cos\theta_\ell
    \end{pmatrix}_0
  =
    \begin{pmatrix}
      m^2_{\tilde \ell_1} & 0 \\
      0 & m^2_{\tilde \ell_2}
    \end{pmatrix}_0
  ~, \label{3.3}
\end{equation}
}while the mass of the $\tilde \nu$ is given by
{\small $(m^2_{\tilde \nu_\ell})_0 =~ (\widetilde m^2_{\tilde \nu_{\ell L}}+ {\frac{1}{2}} 
                        M_Z^2\cos 2\beta)_0 $} . 
We assume the SU(2)$_L$ conditions on their left-handed soft SUSY-breaking mass terms,
{\small $\widetilde m^2_{\tilde e_L}=\widetilde m^2_{\tilde \nu_{eL}}$},
{\small $\widetilde m^2_{\tilde \mu_L}=\widetilde m^2_{\tilde \nu_{\mu L}}$} and
{\small $\widetilde m^2_{\tilde \tau_L}=\widetilde m^2_{\tilde \nu_{\tau L}}$}, which lead to
the following condition, 
{\small
\begin{equation}
   m^2_{\tilde \nu_\ell}
   = \cos^2\theta_\ell m^2_{\tilde \ell_1} 
     +\sin^2\theta_\ell m^2_{\tilde \ell_2} -m_\ell^2+ M_W^2 \cos2\beta ~. \label{3.4}
\end{equation} 
}This relation is valid among the bare quantities as well as among the renormalized quantities.

In the slepton sector, there are three mass renormalization constants, five wavefunction
renormalization constants, and one mixing angle renormalization constants for each generation.
They are
{\small $~\delta m_{\tilde \ell_1},~ \delta m_{\tilde \ell_2},~ \delta m_{\tilde \nu_\ell},~
  \delta Z_{\tilde \ell_i \tilde \ell_j}~(i,j=1,2),~ \delta Z_{\tilde \nu_\ell},~
  \delta \theta_\ell~$}.
We introduce the wavefunction renormalization constants in the unmixed fields
{\small $\tilde \ell_L$} and {\small $\tilde \ell_R$} for each charged slepton. 
{\small
\begin{equation}
    \begin{pmatrix}
      \tilde \ell_L \\
      \tilde \ell_R
    \end{pmatrix}_0
   =
    \begin{pmatrix}
      Z_L^{1/2} & 0 \\
      0 & Z^{1/2}_R
    \end{pmatrix}
    \begin{pmatrix}
      \tilde \ell_L \\
      \tilde \ell_R
    \end{pmatrix}
  ~, \label{3.5}
\end{equation} 
}which can be also written as
{\small
\begin{eqnarray}
    \begin{pmatrix}
       \tilde \ell_L \\
       \tilde \ell_R
    \end{pmatrix}_0
  &=&
    \begin{pmatrix}
       \cos\theta_\ell & -\sin\theta_\ell\\
       \sin\theta_\ell & \cos\theta_\ell
    \end{pmatrix}_0
    \begin{pmatrix}
       \tilde \ell_1 \\
       \tilde \ell_2
    \end{pmatrix}_0   \nonumber\\
  &=&
    \begin{pmatrix}
       \cos\theta_\ell & -\sin\theta_\ell\\
       \sin\theta_\ell & \cos\theta_\ell
    \end{pmatrix}_0
    \begin{pmatrix}
       Z_{11}^{1/2} & Z_{11}^{1/2} \\
       Z_{21}^{1/2} & Z_{22}^{1/2}
    \end{pmatrix}
    \begin{pmatrix}
       \tilde \ell_1 \\
       \tilde \ell_2
    \end{pmatrix}
  ~.  \label{3.6}
\end{eqnarray}
}The first equation of (\ref{3.6}) means, in particular,
{\small
\begin{equation}
    \begin{pmatrix}
       \tilde \ell_L \\
       \tilde \ell_R
    \end{pmatrix}
   =
    \begin{pmatrix}
       \cos\theta_\ell & -\sin\theta_\ell\\
       \sin\theta_\ell & \cos\theta_\ell
    \end{pmatrix}
    \begin{pmatrix}
       \tilde \ell_1 \\
       \tilde \ell_2
    \end{pmatrix}
  ~, \label{3.7}
\end{equation}
}and inserting this expression in (\ref{3.5}), we find
{\small
\begin{equation}
    \begin{pmatrix}
       \tilde \ell_L \\
       \tilde \ell_R
    \end{pmatrix}_0
   =
    \begin{pmatrix}
      Z_L^{1/2} & 0 \\
      0 & Z^{1/2}_R
    \end{pmatrix}
    \begin{pmatrix}
       \cos\theta_\ell & -\sin\theta_\ell\\
       \sin\theta_\ell & \cos\theta_\ell
    \end{pmatrix}
    \begin{pmatrix}
       \tilde \ell_1 \\
       \tilde \ell_2
    \end{pmatrix}
  ~. \label{3.8}
\end{equation}
}Therefore, four wavefunction renormalization constants,
{\small $Z^{1/2}_{11}$}, {\small $Z^{1/2}_{12}$}, {\small $Z^{1/2}_{21}$},
{\small $Z^{1/2}_{22}$} can be expressed in terms of three independent
renormailzation constants, {\small $Z^{1/2}_L$}, {\small $Z^{1/2}_R$} and
{\small $\delta\theta_\ell$}, for each charged slepton.

We have adopted the following renormalization conditions in our
paper on the chargino pair-production and decays \cite{Fujimoto:2007bn}.
\begin{itemize}
\item the on mass-shell conditions for all the three sleptons in each generation
\item the residue conditions for all the three sleptons in each generation
\item the decoupling conditions for the on-shell $\tilde \ell_i$
with $\tilde \ell_j ,(j\ne i, \tilde \ell = \tilde e, \tilde \mu, \tilde \tau)$
\item SU(2) relation for $\delta\theta_\ell$
\end{itemize}
which lead to the following expressions for the renormalization 
constants.
{\small
\begin{eqnarray}
     \delta m_{\tilde \ell}^2 &=& -Re \Sigma_{\tilde \ell \tilde \ell}(m_{\tilde \ell}^2)~,~~~ 
       \tilde \ell = \tilde e_1, \tilde e_2, \tilde \nu_e,
                     \tilde \mu_1, \tilde \mu_2, \tilde \nu_\mu,
                     \tilde \tau_1, \tilde \tau_2, \tilde \nu_\tau ~,  \label{3.10} \\
   \delta Z_{\tilde \ell \tilde \ell} &=&\Sigma^\prime(m_{\tilde \ell}^2)~,~~~
       \tilde \ell = \tilde e_1, \tilde e_2, \tilde \mu_1, \tilde \mu_2,
                     \tilde \tau_1, \tilde \tau_2 ~,       \label{3.11} \\
   \delta Z_{\tilde \nu_\ell}&=&\Sigma^\prime(m_{\tilde \nu_\ell}^2) ,~~~
              \ell = e, \mu, \tau ~,    \label{3.12} \\
   {\frac{1}{2}}\delta Z_{\tilde \ell_i \tilde \ell_j} 
      &=& - {\frac{\Sigma_{\tilde \ell_i \tilde \ell_j}(m_{\tilde \ell_j}^2)}
        {m_{\tilde \ell_i}^2-m_{\tilde \ell_j}^2}}~, ~~~i\ne j,
        ~~\tilde \ell = \tilde e, \tilde \mu, \tilde \tau ~, \label{3.13}  \\
    \delta\theta_\ell
    &=& {\frac{ \delta m_{\tilde \nu_\ell} -\delta(M_W^2\cos2\beta-m_\ell^2)
                         -\cos^2\theta_\ell \delta m^2_{\tilde \ell_1}
                         -\sin^2\theta_\ell \delta m^2_{\tilde \ell_2} }
       { \sin 2\theta_\ell (m^2_{\tilde \ell_2}-m^2_{\tilde \ell1}) }} ~,~~~
          \ell = e, \mu, \tau ~. \label{3.14}
\end{eqnarray}
}

\section{Extension of non-linear gauge formalism}
We can extend the NLG functions (\ref{L2}) and (\ref{L3}) by including bilinear forms
of sleptons with new NLG parameters $\tilde{c}$'s~\cite{Jimbo:2010} as follows:
{\small
\begin{eqnarray}
 F_{W^+} &=& (\partial_\mu + ie\tilde\alpha A_\mu  
                  + ig c_W\tilde\beta Z_\mu) W^{+\mu}
            + i\xi_W {\frac{g}{2}}(v + \tilde\delta_H H^0 + \tilde\delta_h h^0 
              + i\tilde\kappa G^0)G^+ \nonumber \\
  &+& i\xi_W g\Bigl[
       \sum_{i=1,2} \left\{
           \tilde c^e_i(\tilde e_i^* \tilde \nu_e)
          +\tilde c^\mu_i(\tilde \mu_i^* \tilde \nu_\mu)
          +\tilde c^\tau_i(\tilde \tau_i^* \tilde \nu_\tau) \right\} \Bigr]~~, \label{S1} \\
 F_Z &=& \partial_\mu Z^\mu+\xi_Z{\frac{g_Z}{2}}(v +\tilde\epsilon_H H^0
            +\tilde\epsilon_h h^0)G^0 \nonumber \\
  &+& \xi_Zg_Z \Bigl[
           \tilde c^{\nu_e\nu_e}(\tilde\nu_e^*\tilde \nu_e)
          +\tilde c^{\nu_\mu\nu_\mu}(\tilde\nu_\mu^*\tilde \nu_\mu)
          +\tilde c^{\nu_\tau\nu_\tau}(\tilde\nu_\tau^*\tilde \nu_\tau) \nonumber \\
  &&      + \sum_{i,j=1,2} \left\{
           \tilde c^{ee}_{ij}(\tilde e_i^*\tilde e_j)
          +\tilde c^{\mu\mu}_{ij}(\tilde \mu_i^*\tilde \mu_j)
          +\tilde c^{\tau\tau}_{ij}(\tilde \tau_i^*\tilde \tau_j) \right\} \Bigr]~~, \label{S2}
\end{eqnarray}
}while {\small $F_{W^-}$} is hermitian conjugate to {\small $F_{W^+}$}.

\begin{figure}[htb]
\begin{tabular}{ccc}
\begin{minipage}{0.2\columnwidth}
\begin{center}
\includegraphics[width=\columnwidth]{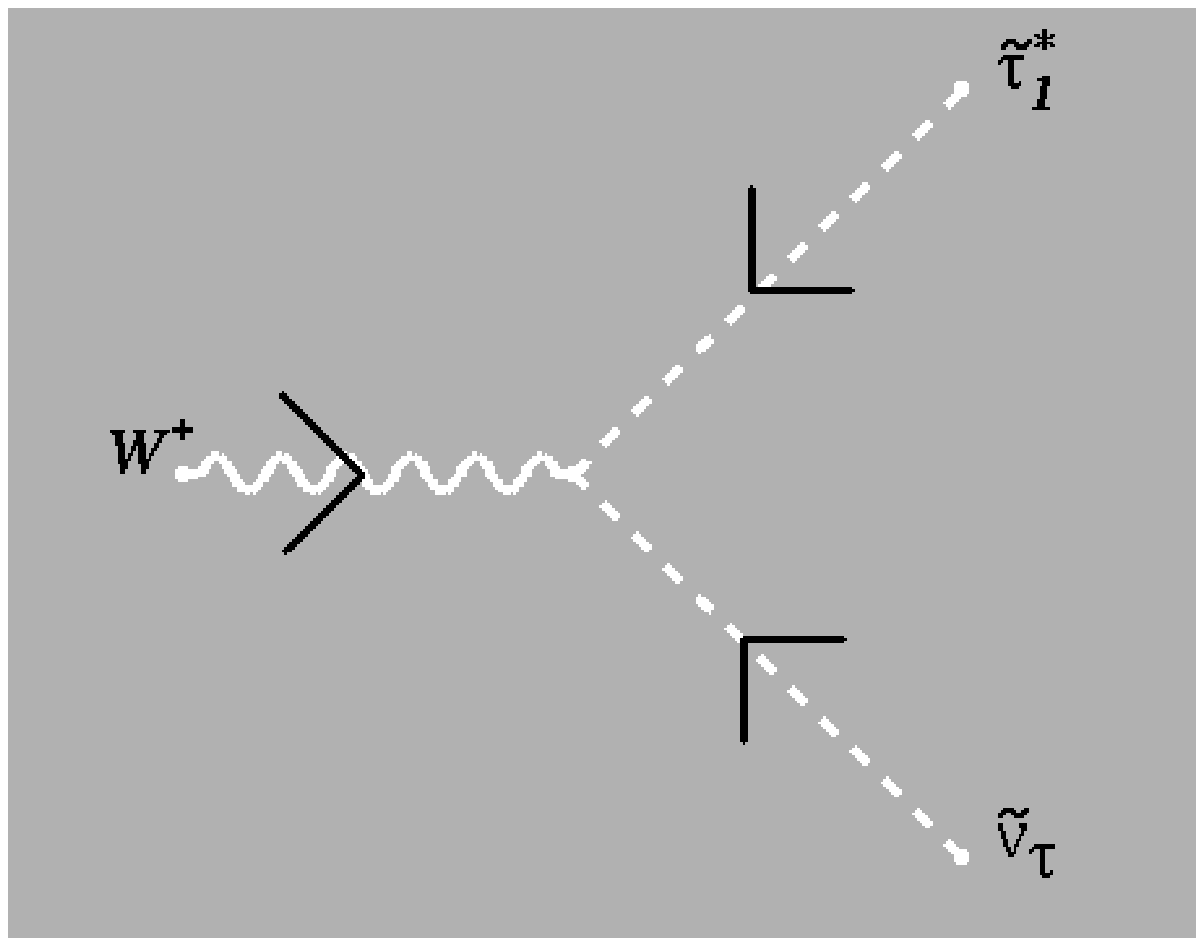}
\label{Fig:FeynRule1}
\end{center}
\end{minipage}
\begin{minipage}{0.02\columnwidth}
\begin{center}
\end{center}
\end{minipage}
\begin{minipage}{0.6\columnwidth}
$\underline{W^+ - \widetilde{\nu_\tau} - \widetilde{\tau_1}}$
{\small
\begin{eqnarray}
i \frac{g}{\sqrt{2}} \cos\theta_\tau (p^{\tilde\nu}_\mu - p^{\tilde\tau}_\mu)
-ig \tilde c^\tau_1 (p^{\tilde\nu}_\mu + p^{\tilde\tau}_\mu) & \nonumber
\end{eqnarray}
}
\end{minipage} \\
\begin{minipage}{0.2\columnwidth}
\begin{center}
\includegraphics[width=\columnwidth]{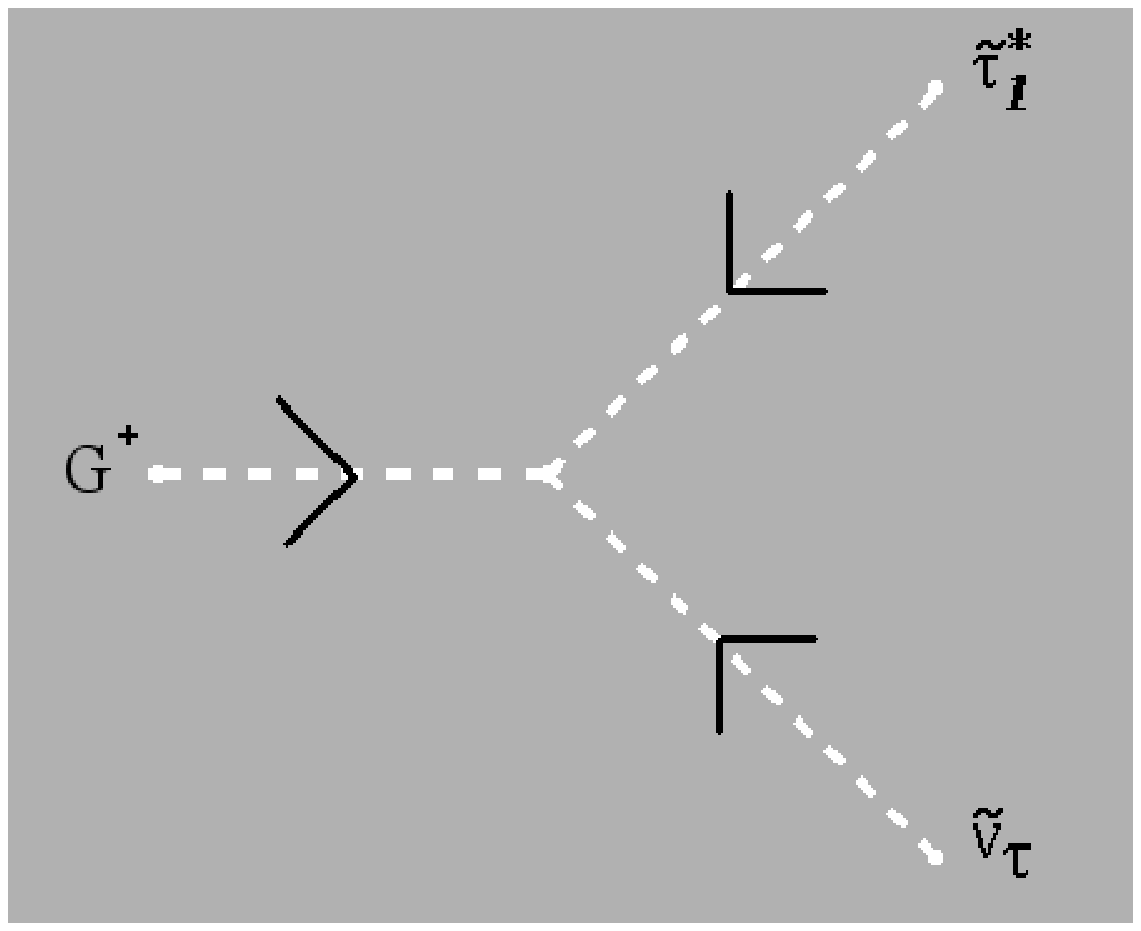}
\label{Fig:FeynRule2}
\end{center}
\end{minipage}
\begin{minipage}{0.02\columnwidth}
\begin{center}
\end{center}
\end{minipage}
\begin{minipage}{0.6\columnwidth}
$\underline{G^+ - \widetilde{\nu_\tau} - \widetilde{\tau_1}}$
{\small
\begin{eqnarray}
i \frac{g}{\sqrt{2} M_W} \left[ \cos\theta_\tau (M_W^2\cos 2\beta -m_\tau^2)
+\sin\theta_\tau m_\tau (A_\tau +\mu\tan\beta) \right]
-ig M_W \tilde c^\tau_1 & \nonumber
\end{eqnarray}
}
\end{minipage} \\
\begin{minipage}{0.2\columnwidth}
\begin{center}
\includegraphics[width=\columnwidth]{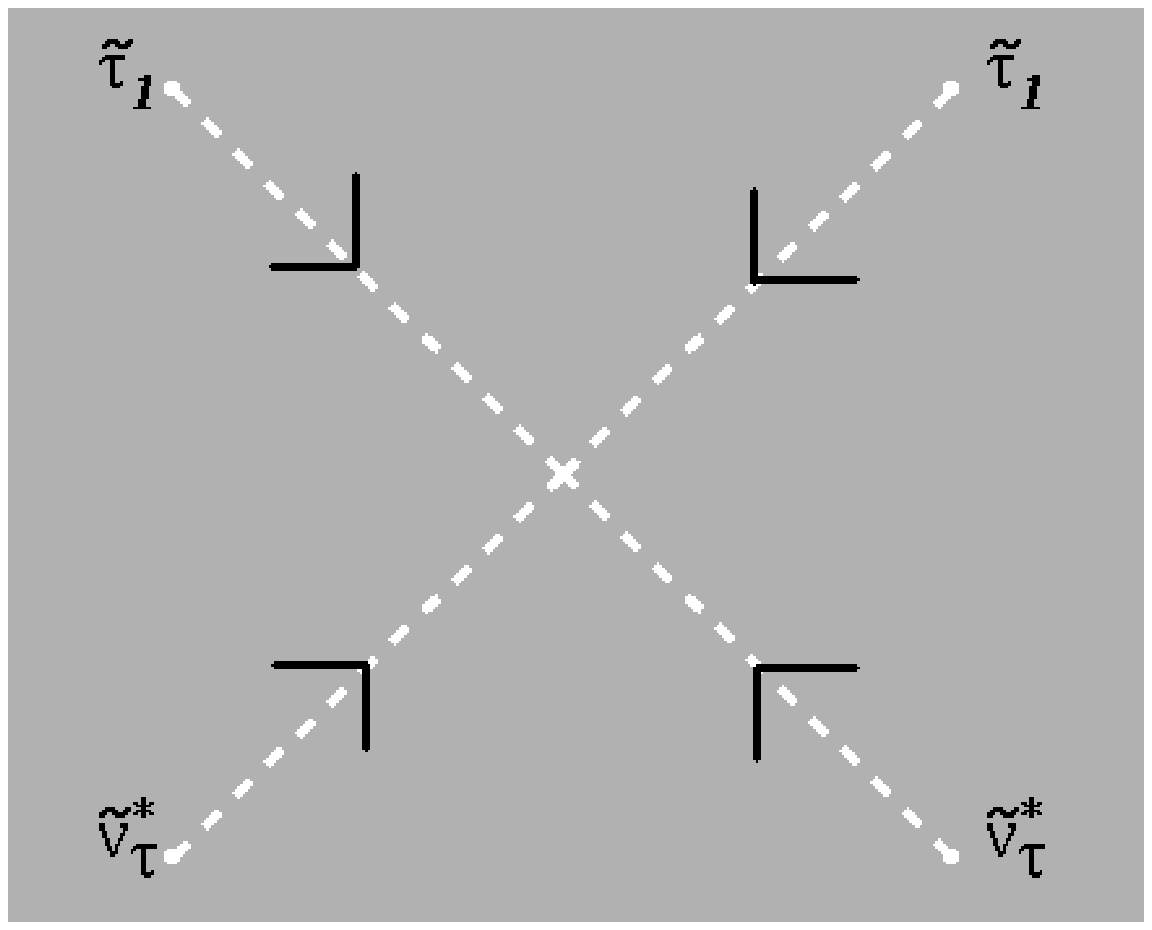}
\label{Fig:FeynRule3}
\end{center}
\end{minipage}
\begin{minipage}{0.02\columnwidth}
\begin{center}
\end{center}
\end{minipage}
\begin{minipage}{0.6\columnwidth}
$\underline{\widetilde{\nu_\tau} - \widetilde{\tau_1} - \widetilde{\nu_\tau} - \widetilde{\tau_1}}$
{\small
\begin{eqnarray}
-i \frac{g_Z^2}{4} \cos^2 \theta_\tau +\frac{i}{2} \sin^2 \theta_\tau
 \left(g_Z^2 \sin\theta_W -{\frac{g^2 m_\tau^2}{M_W^2\cos^2\beta}} \right)
-ig M_W \tilde c^\tau_1 & \nonumber
\end{eqnarray}
}
\end{minipage}
\end{tabular}
\caption{Feynman rules including the NLG couplings}\label{Fig:FeynRules}
\end{figure}

In this paper, we focus on the NLG parameter $\tilde c^\tau_1$ in the NLG function {\small $F_{W^\pm}$},
and set {\small $\xi_W = 1$} in order to avoid the instability in the one-loop calculations.
The Feynman rules of vertices in the linear gauge are modified by introducing the NLG parameter
$\tilde c^\tau_1$ as shown in Figure \ref{Fig:FeynRules}.

The gauge invariance of NLG in the one-loop calculations is guaranteed
by the BRST transformation, which leads to the introducion of the Faddeev-Poppov
ghosts, $\omega_\pm, \omega_Z$ and $\omega_\gamma$, and
anti-ghosts, $\bar \omega_\pm, \bar \omega_Z$ and $\bar \omega_\gamma$.
The corresponding ghost lagrangian to the NLG parameter $\tilde c^\tau_1$ is given as follows:
{\small
\begin{eqnarray}
   {\cal L}_{ghost} &=&
  -i\xi_W g \tilde c^\tau_1 \bar\omega_+
          \left({\frac{i}{2}}\right)
          \Bigl[ \sqrt 2g \cos\theta_\tau \omega_+\tilde\nu_\tau^*
                -\cos\theta_\tau g_Z\omega_Z(\cos\theta_\tau \tilde\tau_1^*-\sin\theta_\tau \tilde\tau_2^*)
          \nonumber \\ 
   && \qquad\qquad\qquad\qquad  -2e\omega_\gamma(\cos\theta_\tau \tilde\tau_1^*-\sin\theta_\tau \tilde\tau_2^*)\Bigr]
           \tilde\nu_\tau \nonumber \\
   && -i\xi_W g \tilde c^\tau_1 \bar\omega_+ \left({\frac{i}{2}}\right)
        (2g_Z \sin\theta_W^2 \omega_z -2e\omega_\gamma)
         (\tilde\tau_1^*)\tilde\nu_\tau   \nonumber \\
   && -i\xi_W g \tilde c^\tau_1 \bar\omega_+\tilde\tau_1^*
          \left({\frac{-i}{2}}\right)
          \Bigl[\sqrt 2g\omega_+(\cos\theta_\tau \tilde\tau_1-\sin\theta_\tau \tilde\tau_2) 
                + g_z\omega_Z\tilde\nu_\tau  \Bigr]
   ~ + (h.c.) ~.
\label{gl}
\end{eqnarray}
}
\begin{figure}[htb]
\centerline{\includegraphics[width=0.95\columnwidth]{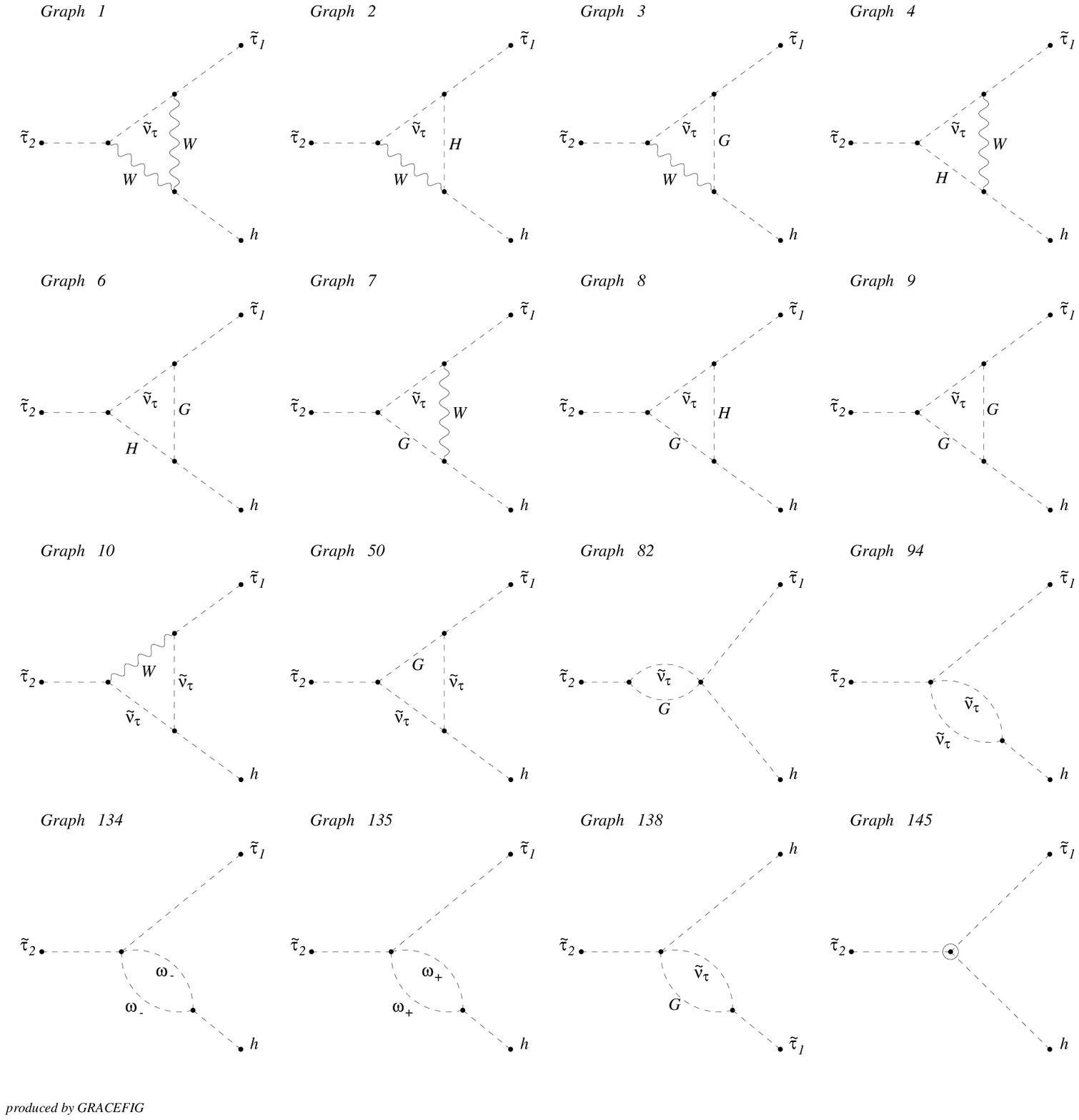}}
\caption{Typical Feynman diagrams of $\tilde\tau_2\to \tilde\tau_1 + h^0$}
\label{Fig:FeynDecay}
\end{figure}

\section{Numerical tests}
We have calculated the slepton decay widths in one-loop order including
the NLG gauge fixing functions (\ref{S1}).  Here we present the results of
the process, $\tilde\tau_2\to \tilde\tau_1 + h^0$.  Typical Feynman diagrams of this
process concerned in the NLG coupling $\tilde c^\tau_1$ is shown in Figure \ref{Fig:FeynDecay}.
We have investigated coefficients of the zeroth power to fourth power of
$\tilde c^\tau_1$ in the UV part and UV finite part.  Table \ref{tab:testsNLG1}
shows nemerical results, in which we have used SUSY parameters as {\small $M_2 = 400$} GeV,
{\small $\mu = - 100$} GeV, {\small $\tan\beta = 30$}, {\small $m_{\tilde\tau_1} = 495.84$} GeV,
{\small $m_{\tilde\tau_2} = 608.23$} GeV and {\small $\theta_\tau = 0.74\pi$}.  Then we have
confirmed the NLG invariance of vertices for the two-body decays in one-loop order.

\begin{figure}[htb]
\centerline{\includegraphics[width=0.9\columnwidth]{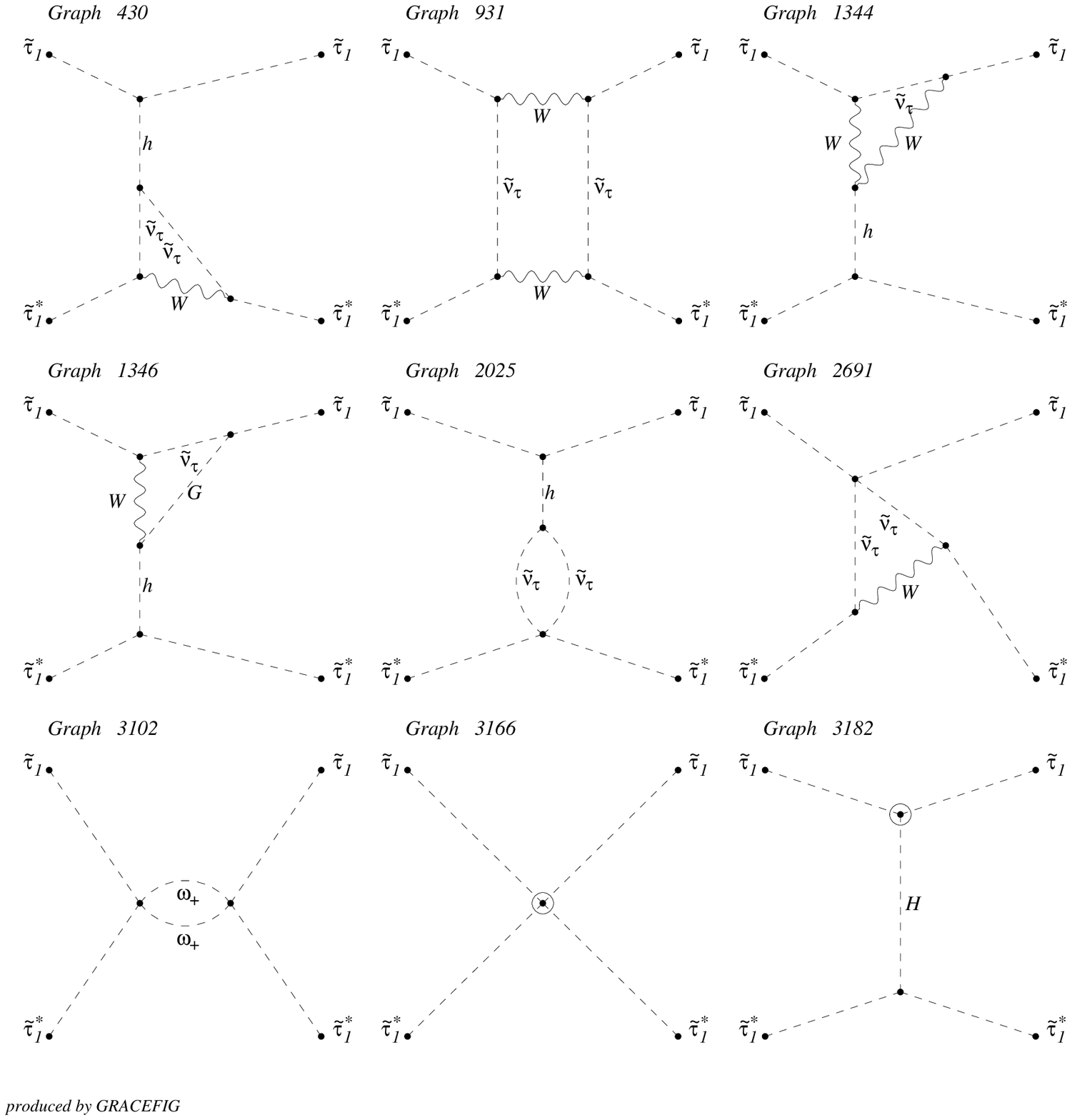}}
\caption{Typical Feynman diagrams of $\tilde\tau_1 + \tilde\tau_1^* \to \tilde\tau_1 + \tilde\tau_1^*$}
\label{Fig:FeynScatt}
\end{figure}

We have also calculated cross sections of the scattering processes systematically
to test the NLG invariance of up to four-point vertices in one-loop order.
Here we present the results of the process,
$\tilde\tau_1 + \tilde\tau_1^* \to \tilde\tau_1 + \tilde\tau_1^*$ .  Typical Feynman
diagrams of this process concerned in the NLG coupling $\tilde c^\tau_1$ is shown
in Figure \ref{Fig:FeynScatt}.
Table \ref{tab:testsNLG2} shows numerical results.

\section{Summary}
We have developed the program package {\tt GRACE/SUSY-loop} for the MSSM amplitudes
in one-loop order, and extended the non-linear gauge formalism applied to {\tt GRACE/SUSY-loop}
by introducing the gauge fixing terms of bilinear forms of sleptons.  Then we have confirmed
the NLG invariance of the MSSM amplitudes in one-loop order for decay processes and
scattering processes using {\tt GRACE/SUSY-loop}.

\section*{Acknowledgments}

This work is partially supported by Grant-in-Aid for Scientific Research(B) (20340063) and
Grant-in-Aid for Scientific Research on Innovative Areas (21105513).


\begin{footnotesize}

\end{footnotesize}


\begin{table}[htb]
\centerline{\scriptsize
\begin{tabular}{|r||r|r|r||r|r|} \hline
 graph & \multicolumn{3}{c||}{UV part} & \multicolumn{2}{c|}{Finite part} 
\\ \hline
\rule{0mm}{9pt}
  & $(\tilde c^\tau_1)^0$ & $(\tilde c^\tau_1)^1$ & $(\tilde c^\tau_1)^2$ & 
 $(\tilde c^\tau_1)^1$ & $(\tilde c^\tau_1)^2$ \\ \hline
 & \multicolumn{5}{c|}{Virtual} \\ \hline
1 & {\tt -2.864048E+02} &{\tt  8.100752E+02} &{\tt -5.728097E+02} &
    {\tt -6.902655E+00} &{\tt  6.310783E+00} \\
2 & {\tt  2.640140E+02} &{\tt -3.733722E+02} &{\tt  0.000000E+00} &
    {\tt  4.340137E+00} &{\tt  0.000000E+00} \\
3 & {\tt  1.281803E+03} &{\tt -2.015262E+03} &{\tt  2.864048E+02} &
    {\tt  2.219785E+01} &{\tt -3.032220E+00} \\
4 & {\tt -2.459420E+02} &{\tt  3.478145E+02} &{\tt  0.000000E+00} &
    {\tt -4.165453E+00} &{\tt  0.000000E+00} \\
6 & {\tt  0.000000E+00} &{\tt  0.000000E+00} &{\tt  0.000000E+00} &
    {\tt  2.651421E-01} &{\tt  0.000000E+00} \\
7 & {\tt -1.470710E+03} &{\tt  1.877379E+03} &{\tt  2.864048E+02} &
    {\tt -2.268075E+01} &{\tt -3.192840E+00} \\
8 & {\tt  0.000000E+00} &{\tt  0.000000E+00} &{\tt  0.000000E+00} &
    {\tt -3.626202E-01} &{\tt  0.000000E+00} \\
9 & {\tt  0.000000E+00} &{\tt  0.000000E+00} &{\tt  0.000000E+00} &
    {\tt -7.996177E-02} &{\tt -8.572352E-02} \\
10& {\tt  1.839419E+02} &{\tt -5.202662E+02} &{\tt  3.678837E+02} &
    {\tt  5.671871E+00} &{\tt -4.659912E+00} \\
50& {\tt  0.000000E+00} &{\tt  0.000000E+00} &{\tt  0.000000E+00} &
    {\tt  1.638727E-02} &{\tt  1.756807E-02} \\
82& {\tt  1.467649E+03} &{\tt  2.020973E+02} &{\tt  0.000000E+00} &
    {\tt -2.152933E+00} &{\tt  0.000000E+00} \\
94& {\tt -9.034420E+01} &{\tt  0.000000E+00} &{\tt -3.678837E+02} &
    {\tt  0.000000E+00} &{\tt  4.642344E+00} \\
134&{\tt  0.000000E+00} &{\tt -2.025188E+02} &{\tt  0.000000E+00} &
    {\tt  1.726551E+00} &{\tt  0.000000E+00} \\
135&{\tt  0.000000E+00} &{\tt -2.025188E+02} &{\tt  0.000000E+00} &
    {\tt  1.726551E+00} &{\tt  0.000000E+00} \\
138&{\tt -1.279135E+03} &{\tt  2.020973E+02} &{\tt  0.000000E+00} &
    {\tt -2.266272E+00} &{\tt  0.000000E+00} \\ \hline
 & \multicolumn{5}{c|}{Counter Term} \\ \hline
145&{\tt  5.330665E+02} &{\tt -1.255253E+02} &{\tt  0.000000E+00} &
    {\tt  2.666161E+00} &{\tt  0.000000E+00} \\ \hline \hline
 &\multicolumn{5}{c|}{Total} \\ \hline
 &  {\tt -6.032040E-20} &{\tt  5.714153E-27} &{\tt  1.019217E-29} &
    {\tt -3.995728E-23} &{\tt -2.514610E-31} \\ \hline
\end{tabular}
}
\caption{Test for NLG invariance of $\tilde\tau_2\to \tilde\tau_1 + h^0$}
\label{tab:testsNLG1}
\end{table}
\begin{table}[htb]
\centerline{\scriptsize
\begin{tabular}{|r||r|r|r|r|r|} \hline
 graph & \multicolumn{5}{c|}{UV part} \\ \hline
\rule{0mm}{9pt}
  & $(\tilde c^\tau_1)^0$ & $(\tilde c^\tau_1)^1$ & $(\tilde c^\tau_1)^2$ & 
 $(\tilde c^\tau_1)^3$ & $(\tilde c^\tau_1)^4$ \\ \hline
 & \multicolumn{5}{c|}{Virtual} \\ \hline
430	& {\tt -5.230496E-02} & {\tt -2.130068E-01 } & {\tt -2.168624E-01} &
    {\tt 0.000000E+00} &{\tt 0.000000E+00} \\
931	& {\tt -9.672486E-03} & {\tt -7.878050E-02} & {\tt -2.406194E-01} &
    {\tt -3.266330E-01} & {\tt -1.662726E-01} \\
1344 & {\tt 8.144091E-02} & {\tt 3.316601E-01} & {\tt 3.376633E-01} &
    {\tt 0.000000E+00} & {\tt 0.000000E+00} \\
1346 & {\tt -3.644882E-01} & {\tt -8.250864E-01} & {\tt -1.688317E-01} &
    {\tt 0.000000E+00} & {\tt 0.000000E+00} \\
2025 & {\tt 4.225897E-02} & {\tt 0.000000E+00} & {\tt 2.168624E-01} &
    {\tt 0.000000E+00} & {\tt 0.000000E+00} \\
2691 & {\tt 7.814734E-03} & {\tt 3.182474E-02} & {\tt 7.250402E-02} &
    {\tt 1.633165E-01} & {\tt 1.662726E-01} \\
3102 & {\tt 0.000000E+00} & {\tt 0.000000E+00} & {\tt 4.010324E-02} &
    {\tt 0.000000E+00} & {\tt 0.000000E+00} \\ \hline
 & \multicolumn{5}{c|}{Counter Term} \\ \hline
3166 & {\tt 6.074190E-02} & {\tt -1.272990E-01} & {\tt 0.000000E+00} &
    {\tt 0.000000E+00} & {\tt 0.000000E+00} \\
3182 & {\tt 1.777580E+02} & {\tt 1.914021E+02} & {\tt 0.000000E+00} &
    {\tt 0.000000E+00} & {\tt 0.000000E+00} \\ \hline \hline
 &\multicolumn{5}{c|}{Total} \\ \hline
 &  {\tt -1.108258E-21} & {\tt -1.813692E-25} & {\tt -9.099667E-29} &
    {\tt 2.117843E-31} & {\tt 4.801768E-38} \\ \hline \hline
 & \multicolumn{5}{c|}{Finite part} \\ \hline
 & \multicolumn{5}{c|}{Virtual} \\ \hline
430	&  & {\tt 2.452794E-03} & {\tt 2.745929E-03} &
    {\tt 0.000000E+00} & {\tt 0.000000E+00} \\
931 &  & {\tt 9.074788E-04} & {\tt 3.123851E-03} &
    {\tt 3.760592E-03} & {\tt 2.111032E-03} \\
1344 &  & {\tt -3.295736E-03} & {\tt -3.852588E-03} &
    {\tt 0.000000E+00} & {\tt 0.000000E+00} \\
1346 &  & {\tt 9.344057E-03} & {\tt 1.939682E-03} &
    {\tt 0.000000E+00} & {\tt 0.000000E+00} \\
2025 &  & {\tt 0.000000E+00} & {\tt -2.738628E-03} &
    {\tt 0.000000E+00} & {\tt 0.000000E+00} \\
2691 &  & {\tt -3.664649E-04} & {\tt -1.035375E-03} &
    {\tt -1.880605E-03} & {\tt -2.105357E-03} \\
3102 &  & {\tt 0.000000E+00} & {\tt -4.706958E-04} &
    {\tt 0.000000E+00} & {\tt 0.000000E+00} \\ \hline
 & \multicolumn{5}{c|}{Counter Term} \\ \hline
3166 &  & {\tt 1.427501E-03} & {\tt 0.000000E+00} &
    {\tt 0.000000E+00} & {\tt 0.000000E+00} \\
3182 &  & {\tt -2.146339E+00} & {\tt 0.000000E+00} &
    {\tt 0.000000E+00} & {\tt 0.000000E+00} \\ \hline \hline
 &\multicolumn{5}{c|}{Total} \\ \hline
 &   & {\tt 3.341261E-27} & {\tt -3.304929E-29} &
   {\tt -4.964405E-29} & {\tt -2.529046E-29} \\ \hline
\end{tabular}
}\caption{Test for NLG invariance of $\tilde\tau_1 + \tilde\tau_1^* \to \tilde\tau_1 + \tilde\tau_1^*$}
\label{tab:testsNLG2}
\end{table}


\begin{thebibliography}{10}

\bibitem{Fujimoto:2007bn}
J.~Fujimoto, T.~Ishikawa, Y.~Kurihara, M.~Jimbo, T.~Kon and M.~Kuroda, Phys. Rev.
{\bf D75} 113002 (2007).
\bibitem{Iizuka:2010bh}
K.~Iizuka, T.~Kon, K.~Kato, T.~Ishikawa, Y.~Kurihara, M.~Jimbo and M.~Kuroda,
PoS(RADCOR2009)068, (arXiv:1001.2800 [hep-ph]) (2010).
\bibitem{Jimbo:2010}
M.~Jimbo, K.~Iizuka, T.~Ishikawa, K.~Kato, T.~Kon, Y.~Kurihara, M. Kuroda,
To appear in {\it Proceedings of International Linear Collider Workshop 2010 (LCWS10 \& ILC10),
Beijing, China, 26-30 Mar 2010}, (arXiv:1006.3491 [hep-ph]) (2010).
\bibitem{Fujikawa:1973} K.~Fujikawa, Phys. Rev {\bf D7} 393 (1973).
\bibitem{Gavela:1981} M.B.~Gavela, G.~Girardi, C.~Malleville and P.~Sorba,
Nucl. Phys. {\bf B193} 257 (1981).
\bibitem{Haber:1988} H.E.~Haber and D.~Wyler, SCIPP-88/19 (1988).
\bibitem{Capdequi:1990} M.~Capdequi-Peyran\`{e}re, H.E.~Haber, P.~Irulegui, PM-90-06, SCIPP-90-03, 
in {\it Rigorous Methods in Particle Physics}, Springer Tract. Phys. {\bf 119} (1990).
\bibitem{Boudjema:1996} F.~Boudjema and A.~Chopin, Z. Phys. {\bf C73} 85 (1996).
\bibitem{Kato:2006}
G.~B\'elanger, F.~Boudjema, J.~Fujimoto, T.~Ishikawa, T.~Kaneko,
K.~Kato, Y.~Shimizu, Phys. Rept. {\bf 430} 117 (2006).
\bibitem{Fujimoto:2006km}
J.~Fujimoto, T.~Ishikawa, M.~Jimbo, T.~Kaneko, T.~Kon, Y.~Kurihara,
M.~Kuroda and Y.~Shimizu, Nucl. Phys. Proc. Suppl. {\bf 157} 157 (2006).
\bibitem{Yuasa:1999rg}
F.~Yuasa, J.~Fujimoto, T.~Ishikawa, M.~Jimbo, T.~Kaneko, K.~Kato,
S.~Kawabata, T.~Kon, Y.~Kurihara, M.~Kuroda, N.~Nakazawa, Y.~Shimizu
and H.~Tanaka, Prog. Theor. Phys. Suppl. {\bf 138} 18 (2000); \\
{\texttt http://minami-home.kek.jp/}~.
\bibitem{Baro:arXiv0906.1665}
N.~Baro and F.~Boudjema, Phys. Rev. {\bf D80} 076010 (2009).
\bibitem{Hahn:2000jm}
T.~Hahn, Nucl. Phys. Proc. Suppl. {\bf 89} 231 (2000);
Comput. Phys. Commun. {\bf 140} 418 (2001).
\bibitem{Chankowski:1994}
P.H.~Chankowski, S.~Pokorski and J.~Rosiek, Nucl. Phys. {\bf B423} 437 (1994).
\bibitem{Yamada:1991}
A.~Yamada, Phys. Lett. {\bf B263} 233 (1991); Z. Phys. {\bf C61} 247 (1994).
\bibitem{Dabelstein:1995}
A.~Dabelstein, Z. Phys. {\bf C67} 495 (1995).
\bibitem{Hollik:2002}
W.~Hollik, E.~Kraus, M.~Roth, C.~Rupp, K.~Sibold and D.~St\"{o}ckinger,
Nucl. Phys. {\bf B639} 3 (2002).
\bibitem{Fritzsche:2002}
T.~Fritzsche and W.~Hollik, Eur. Phys. J {\bf C24} 619 (2002).

\end{thebibliography}
\end{document}